\documentclass[aps,pra,twocolumn,superscriptaddress]{revtex4}
\usepackage{amsfonts}
\usepackage{amsmath}
\usepackage{amssymb}
\usepackage{graphicx}
\usepackage[english]{babel}
\usepackage{hyperref}
\usepackage[usenames,dvipsnames]{color}
\usepackage{babel}

\begin{document}

\title{1- and 3-photon dynamical Casimir effects using nonstationary cyclic
qutrit}
\author{H. Dessano}
\affiliation{Institute of Physics, University of Brasilia, 70910-900, Brasilia, Federal
District, Brazil}
\affiliation{Instituto Federal de Bras\'{\i}lia, Campus Recanto das Emas, 72620-100,
Brasilia, Federal District, Brazil}
\author{A. V. Dodonov }
\email{adodonov@fis.unb.br}
\affiliation{Institute of Physics, University of Brasilia, 70910-900, Brasilia, Federal
District, Brazil}
\affiliation{International Centre for Condensed Matter Physics, University of Brasilia,
70910-900, Brasilia, Federal District, Brazil}

\begin{abstract}
We consider the nonstationary circuit QED setup in which a 3-level
artificial atom in the $\Delta$-configuration interacts with a single-mode
cavity field of natural frequency $\omega $. It is demonstrated that when
some atomic energy level(s) undergoes a weak harmonic modulation, photons
can be generated from vacuum via effective 1- and 3-photon transitions,
while the atom remains approximately in the ground state. These phenomena
occur in the dispersive regime when the modulation frequency is accurately
tuned near $\omega $ and $3\omega $, respectively, and the generated field
states exhibit strikingly different statistics from the squeezed vacuum
state attained in standard cavity dynamical Casimir effect.
\end{abstract}

\maketitle

\section{Introduction}

The term \emph{cavity dynamical Casimir effect} (DCE) can be used to denote the
class of phenomena that feature the generation of photons from vacuum in
some cavity due to the resonant external perturbation of the system
parameters, where the cavity serves to produce a resonant enhancement of the
DCE \cite{rev0,rev1,nor,rev2,nation,macri}. These phenomena were originally
studied in the context of electromagnetic resonators with oscillating walls
or containing a macroscopic dielectric medium with time-modulated internal
properties \cite{zer,nikonov,law,lambrecht,soh,maianeto}, but later were
generalized for other bosonic fields, e. g., phononic excitations of ion
chains \cite{ions}, optomechanical systems \cite{opto}, cold atoms \cite%
{tito} and Bose-Einstein condensates \cite{bec1,bec2}. For single-mode
cavities the main resonance occurs near the modulation frequency $2\omega $,
where $\omega $ is the bare cavity frequency, and in the absence of
dissipation the average photon number grows exponentially with time \cite%
{pla,evd}, resulting in the squeezed vacuum state with even photon numbers,
analogously to the phenomenon of parametric amplification \cite%
{rev0,rev2,law}. The cavity DCE was recently implemented experimentally
using a Josephson metamaterial consisting of an array of 250 superconductive
interference devices (SQUIDs) embedded in a microwave cavity whose
electrical length was modulated by an external magnetic flux \cite{meta}.

The concept of cavity DCE has been successfully extended to the area of circuit
Quantum Electrodynamics (circuit QED) \cite%
{jpcs,liberato,zeilinger,JPA}, in which one or several artificial Josephson
atoms strongly interact with a microwave field confined in superconducting
resonators and waveguides \cite{cir1,cir2,cir3,cir4}. The exquisite \emph{in situ} control
over the atomic parameters allows to rapidly modulate the atomic energy
levels and the atom-field coupling strength \cite{majer,ge,ger,ger1,v1,v2,v3}%
, enabling the use of artificial atoms as substitutes of the dielectric
medium with time-dependent properties. From the viewpoint of a toy model \cite{igor}, a
modulated or oscillating dielectric slab can be imagined as a set of atoms
with varying parameters, so ultimately DCE must emerge for a single nonstationary $2$%
-level atom. Indeed, it was shown that for off-resonant qubit(s) undergoing a weak external perturbation, pairs of
photons are generated from vacuum under the modulation frequency $\sim
2\omega $ while the atom(s) remains approximately in the initial state \cite%
{jpcs,JPA,igor,tom}. In this scenario the atom plays the role of both the
source and real-time detector of DCE, since the (small) atomic
transition probability depends on the photon number and in turn affects the
photon generation pattern \cite{PLAI,jpcs,diego}. Moreover, the rich nonharmonic
spectrum of the composite atom--field system permits the implementation of
other phenomena in the nonstationary regime, such as: sideband transitions
\cite{blais-exp,schuster,side2}, anti-dynamical Casimir effect \cite%
{igor,diego,lucas,juan,werlang}, $n$-photon Rabi model \cite{nr}, generation
of entanglement \cite{entan,etc3}, quantum simulations \cite%
{relativistic,sim,ger1} and dynamical Lamb effect \cite{lamb,etc1}.

Here we explore theoretically the prospects of implementing nontraditional versions of
cavity DCE using 3-level atoms (qutrits) in the cyclic (also
known as $\Delta $-) configuration subject to parametric modulation. In this case all the transitions between the
atomic levels can occur simultaneously via the cavity field \cite%
{cycl1,cycl2,cycl3,cycl4}, so the total number of excitations is not
conserved even upon neglecting the counter-rotating terms (rotating wave
approximation). Although prohibited by the electric-dipole selection rules for usual atoms, the $\Delta $-configuration can be implemented for certain artificial
atoms in circuit QED \cite{cir4} by breaking the inversion symmetry of the
potential energy. Our goal is to find new modulation
frequencies, exclusive of the cyclic qutrits, that induce photon generation
from vacuum without changing appreciably the
atomic state.

We find that for the harmonic modulation of some energy level(s) of a
dispersive cyclic qutrit, photons can be generated from vacuum for the
modulation frequencies $\eta \approx \omega $ and $\eta \approx 3\omega $
while the atom predominantly remains in the ground state. We call these
processes 1- and 3-photon DCE because the photons are generated via
effective 1- and 3-photon transitions between the system dressed-states,
whose rates depend on the product of all the three coupling strengths. We
derive an approximate analytical description of the unitary dynamics and
illustrate the typical system behavior by solving numerically the Schr\"{o}%
dinger equation. In particular, we show that the average photon number and
atomic populations display a collapse-revival behavior, and the photon
number distributions are completely different from the standard (2-photon)
cavity DCE case. Moreover, we solve numerically the Markovian master
equation and demonstrate that in the presence of weak dissipation the
dissipative dynamics resembles the unitary one for initial times, confirming
that our proposal is experimentally feasible.

\section{Physical system}

We consider a single cavity mode of constant frequency $\omega $ that
interacts with a qutrit in the cyclic configuration \cite%
{cir4,cycl1,cycl2,cycl3,cycl4}, so that all the atomic transitions are
allowed via one-photon transitions. The Hamiltonian reads%
\begin{equation}
\hat{H}/\hbar =\omega \hat{n}+\sum_{k=1}^{2}E_{k}(t)\hat{\sigma}%
_{k,k}+\sum_{k=0}^{1}\sum_{l>k}^{2}g_{k,l}(\hat{a}+\hat{a}^{\dagger })(%
\hat{\sigma}_{l,k}+\hat{\sigma}_{k,l}).  \label{H2}
\end{equation}%
$\hat{a}$ ($\hat{a}^{\dagger }$) is the cavity annihilation (creation)
operator and $\hat{n}=\hat{a}^{\dagger }\hat{a}$ is the photon number
operator. The atomic eigenenergies are $E_{0}\equiv 0,E_{1}$ and $E_{2}$,
the corresponding states are $|\mathbf{k}\rangle $ and we defined $\hat{%
\sigma}_{k,j}\equiv |\mathbf{k}\rangle \langle \mathbf{j}|$. The constant
parameters $g_{k,l}$ denote the coupling strengths between the atomic states
$|\mathbf{k}\rangle $ and $|\mathbf{l}\rangle $ mediated by the cavity
field. To emphasize the role of the counter-rotating terms (CRT) we rewrite
(for $l>k$)%
\begin{equation*}
g_{k,l}(\hat{a}+\hat{a}^{\dagger })(\hat{\sigma}_{l,k}+\hat{\sigma}%
_{k,l})\rightarrow g_{k,l}(\hat{a}\hat{\sigma}_{l,k}+c_{k,l}\hat{a}\hat{%
\sigma}_{k,l}+h.c.),
\end{equation*}%
where $c_{k,l}=1$ when the corresponding CRT is taken into account and is
zero otherwise.

Utilizing the tunability of Josephson atoms  \cite{majer,ge,ger,ger1,v1,v2,v3}, we assume that the atomic energy levels can be modulated externally as
\begin{equation*}
E_{k}(t)\equiv E_{k}^{(0)}+\varepsilon _{k}\sin (\eta t+\phi _{k})\quad \text{ for}%
~k=1,2~,
\end{equation*}%
where $\varepsilon _{k}\ll E_{k}^{(0)}$ is the modulation amplitude, $\phi
_{k}$ is the associated phase, $E_{k}^{(0)}$ is the bare energy value and $%
\eta \gtrsim \omega $ is the modulation frequency. We would like to stress
that for weak perturbations our approach can be easily generalized to
multi-tone modulations or simultaneous perturbation of all the parameters in
Hamiltonian (\ref{H2}).

We expand the wavefunction as
\begin{equation}
|\psi (t)\rangle =\sum_{n=0}^{\infty }e^{-it\lambda _{n}}b_{n}(t)\mathcal{F}%
_{n}(t)|\varphi _{n}\rangle  \label{state}
\end{equation}%
\begin{equation*}
\mathcal{F}_{n}(t)=\exp \left\{ \sum_{k=1}^{2}\frac{i\varepsilon _{k}}{\eta }%
\left[ \cos (\eta t+\phi _{k})-1\right] \langle \varphi _{n}|\hat{\sigma}%
_{k,k}|\varphi _{n}\rangle \right\} .
\end{equation*}%
Here $\lambda _{n}$ are the eigenfrequencies of the bare Hamiltonian $\hat{H}%
_{0}\equiv \hat{H}[\varepsilon _{1}=\varepsilon _{2}=0]$ ($n$ increasing
with energy) and $|\varphi _{n}\rangle $ are the corresponding eigenstates
(dressed-states). $b_{n}(t)$ denotes the slowly-varying probability
amplitude of the state $|\varphi _{n}\rangle $ and $\mathcal{F}%
_{n}(t)\approx 1$ is a rapidly oscillating function with a small amplitude.

After substituting Eq. (\ref{state}) into the Schr\"{o}dinger equation, to
the first order in $\varepsilon _{1}$ and $\varepsilon _{2}$ we obtain the
differential equation%
\begin{equation}
\dot{b}_{n}=\sum\limits_{m\neq n}b_{m}\left[ \Theta _{m;n}^{\ast
}e^{it\left( \lambda _{n}-\lambda _{m}-\eta \right) }-\Theta
_{n;m}e^{-it\left( \lambda _{m}-\lambda _{n}-\eta \right) }\right]
\label{dif}
\end{equation}%
that describes transitions between the dressed-states $|\varphi _{n}\rangle $
and $|\varphi _{m}\rangle $ with the transition rate $|\Theta _{n;m}|$, where%
\begin{equation}
\Theta _{n;m}\equiv \frac{1}{2}\sum_{k=1}^{2}\varepsilon _{k}e^{i\phi
_{k}}\langle \varphi _{n}|\hat{\sigma}_{k,k}|\varphi _{m}\rangle .
\label{rate}
\end{equation}%
The transition $|\varphi _{n}\rangle \leftrightarrow |\varphi _{m}\rangle $
occurs when the modulation frequency is resonantly tuned to $\eta
_{r}=|\lambda _{m}-\lambda _{n}|+\Delta \nu $, where $\Delta \nu $ denotes a
small shift \cite{JPA} dependent on $\varepsilon _{1},\varepsilon _{2}$ due
to the rapidly-oscillating terms that were neglected in Eq. (\ref{dif}) (in
this paper we adjust $\Delta \nu $ numerically). By writing the
interaction-picture wavefunction as $|\psi _{I}(t)\rangle
=\sum_{n}b_{n}(t)|\varphi _{n}\rangle $ one can cast Eq. (\ref{dif}) as a
dressed-picture \emph{effective Hamiltonian}%
\begin{equation*}
\hat{H}_{ef}(t)=-i\sum_{n,m\neq n}\Theta _{m;n}|\varphi _{m}\rangle \langle
\varphi _{n}|e^{-it\left( \lambda _{n}-\lambda _{m}-\eta \right) }+h.c.
\end{equation*}

Since we focus on transitions in which the atom is minimally disturbed, we
consider the dispersive regime%
\begin{equation*}
|\Delta _{1}|,|\Delta _{2}|,|\Delta _{1}+\Delta _{2}|\gg \sqrt{n_{\max }}%
\max (g_{k,l})~,
\end{equation*}%
where $n_{\max }$ is the maximum number of the system excitations and the
bare detunings are defined as%
\begin{equation*}
\Delta _{1}\equiv \omega -E_{1}^{(0)},~\Delta _{2}\equiv \omega
-(E_{2}^{(0)}-E_{1}^{(0)}),~\Delta _{3}\equiv \Delta _{1}+\Delta _{2}.
\end{equation*}%
Denoting by $|\zeta _{k}\rangle $ the dressed-states in which the atom is
predominantly in the ground state, from the standard perturbation theory we
find%
\begin{eqnarray}
|\zeta _{k}\rangle &\approx &|\mathbf{0},k\rangle +\frac{c_{0,1}g_{0,1}^{2}%
\sqrt{k(k-1)}}{2\Delta _{1}\omega }|\mathbf{0},k-2\rangle  \label{ds} \\
&&+\frac{g_{0,1}\sqrt{k}}{\Delta _{1}}|\mathbf{1},k-1\rangle -\frac{%
c_{0,1}g_{0,1}\sqrt{k+1}}{2\omega -\Delta _{1}}|\mathbf{1},k+1\rangle  \notag
\\
&&-\frac{c_{1,2}g_{0,1}g_{1,2}k}{\Delta _{1}(2\omega -\Delta _{3})}|\mathbf{2%
},k\rangle +\frac{g_{0,1}g_{1,2}\sqrt{k(k-1)}}{\Delta _{1}\Delta _{3}}|%
\mathbf{2},k-2\rangle  \notag \\
&&-\frac{g_{0,2}\sqrt{k}}{\omega -\Delta _{3}}|\mathbf{2},k-1\rangle -\frac{%
c_{0,2}g_{0,2}\sqrt{k+1}}{3\omega -\Delta _{3}}|\mathbf{2},k+1\rangle  \notag
\end{eqnarray}%
where $|\mathbf{j},k\rangle \equiv |\mathbf{j}\rangle _{atom}\otimes
|k\rangle _{field}$ and $k\geq 0$. The corresponding eigenfrequencies are
(neglecting constant shifts)%
\begin{equation}
\Lambda _{k}\approx \omega _{ef}k+\alpha k^{2}  \label{d}
\end{equation}%
with the effective cavity frequency and the one-photon Kerr nonlinearity, respectively, %
\begin{eqnarray*}
\omega _{ef} &\equiv &\omega +\frac{g_{0,1}^{2}}{\Delta _{1}}\left( 1-\frac{%
g_{1,2}^{2}}{\Delta _{1}\Delta _{3}}\right) -\frac{g_{0,2}^{2}}{\omega
-\Delta _{3}} \\
&&-\frac{c_{0,1}g_{0,1}^{2}}{2\omega -\Delta _{1}}-\frac{c_{0,2}g_{0,2}^{2}}{%
3\omega -\Delta _{3}}
\end{eqnarray*}%
\begin{eqnarray*}
\alpha &\equiv &\frac{g_{0,1}^{2}}{\Delta _{1}^{2}}\left( \frac{g_{1,2}^{2}}{%
\Delta _{3}}-\frac{g_{0,1}^{2}}{\Delta _{1}}+\frac{c_{0,1}g_{0,1}^{2}}{%
2\omega }-\frac{c_{1,2}g_{1,2}^{2}}{2\omega -\Delta _{3}}\right. \\
&&\left. +\frac{g_{0,2}^{2}}{\omega -\Delta _{3}}+\frac{c_{0,1}g_{0,1}^{2}}{%
2\omega -\Delta _{1}}+\frac{c_{0,2}g_{0,2}^{2}}{3\omega -\Delta _{3}}\right)
.
\end{eqnarray*}%
In the Appendix \ref{apen} we present the complete expressions for the eigenstates and
eigenvalues obtained from the 2- and 4-order perturbation theory,
respectively.

\section{1- and 3-photon DCE}

The lowest-order phenomena that occur exclusively for cyclic qutrits depend
on the combination $g_{0,1}g_{1,2}g_{0,2}$, so we define $G^{3}\equiv
g_{0,1}g_{1,2}g_{0,2}/2$. Indeed, for $g_{0,2}=0$ we recover the ladder
configuration, for $g_{0,1}=0$ the $\Lambda $- configuration and for $%
g_{1,2}=0$ the V-configuration. After substituting the dressed-states (%
\ref{ape}) into Eq. (\ref{rate}) we find that one such effect is the
three-photon transition between the states $|\zeta _{k}\rangle $ and $|\zeta
_{k+3}\rangle $. To the lowest order the respective transition rate reads%
\begin{equation}
\Theta _{k;k+3}^{(\zeta )}=G^{3}\sqrt{\frac{(k+3)!}{k!}}\left[ \varepsilon
_{1}q_{1}e^{i\phi _{1}}-\varepsilon _{2}q_{2}e^{i\phi _{2}}\right] ,
\label{x1}
\end{equation}%
where the $k$-independent parameters are%
\begin{eqnarray*}
q_{1} &=&\frac{c_{0,2}}{\Delta _{1}\left( 3\omega -\Delta _{3}\right) \left(
3\omega -\Delta _{1}\right) } \\
&&+\frac{c_{0,1}c_{1,2}}{\left( 2\omega -\Delta _{1}\right) \left( \omega
-\Delta _{3}\right) \left( \omega +\Delta _{1}\right) }
\end{eqnarray*}%
\begin{eqnarray*}
q_{2} &=&\frac{c_{0,2}}{\Delta _{1}\Delta _{3}\left( 3\omega -\Delta
_{3}\right) } \\
&&+\frac{c_{0,1}c_{1,2}}{\left( 2\omega -\Delta _{1}\right) \left( \omega
-\Delta _{3}\right) \left( 4\omega -\Delta _{3}\right) }~.
\end{eqnarray*}%
We see that this effect, corresponding roughly to the transitions $|\mathbf{0%
},k\rangle \leftrightarrow |\mathbf{0},k+3\rangle \leftrightarrow |\mathbf{0}%
,k+6\rangle \leftrightarrow \cdots $, relies on the CRT: either $c_{0,2}$
must be nonzero, or the product $c_{0,1}c_{1,2}$.

The second effect allowed by the cyclic configuration is the one-photon
transition between the states $|\zeta _{k}\rangle $ and $|\zeta
_{k+1}\rangle $, or roughly $|\mathbf{0},k\rangle \leftrightarrow |\mathbf{0}%
,k+1\rangle \leftrightarrow |\mathbf{0},k+2\rangle \leftrightarrow \cdots $.
We obtain to the lowest order%
\begin{equation}
\Theta _{k;k+1}^{(\zeta )}=G^{3}\sqrt{k+1}\left[ \varepsilon
_{1}Q_{1}(k)e^{i\phi _{1}}-\varepsilon _{2}Q_{2}(k)e^{i\phi _{2}}\right] ,
\label{x2}
\end{equation}%
where we defined $k$-dependent functions%
\begin{eqnarray*}
Q_{1}(k) &=&\frac{1}{\Delta _{1}(\omega -\Delta _{1})}\left( \frac{%
c_{1,2}c_{0,2}(k+1)}{3\omega -\Delta _{3}}+\frac{k}{\omega -\Delta _{3}}%
\right) \\
&&-\frac{c_{0,1}c_{0,2}(k+2)}{(2\omega -\Delta _{1})(3\omega -\Delta
_{3})(3\omega -\Delta _{1})} \\
&&-\frac{c_{1,2}k}{\Delta _{1}(\omega -\Delta _{3})(\omega +\Delta _{1})}-%
\frac{c_{0,1}}{(\omega -\Delta _{1})(2\omega -\Delta _{1})} \\
&&\times \left( \frac{c_{1,2}c_{0,2}(k+2)}{3\omega -\Delta _{3}}+\frac{k+1}{%
\omega -\Delta _{3}}\right)
\end{eqnarray*}%
\begin{eqnarray*}
Q_{2}(k) &=&\frac{1}{(2\omega -\Delta _{3})(\omega -\Delta _{3})}\left(
\frac{c_{0,1}(k+1)}{2\omega -\Delta _{1}}-\frac{c_{1,2}k}{\Delta _{1}}\right)
\\
&&+\frac{c_{0,1}c_{1,2}c_{0,2}(k+2)}{(2\omega -\Delta _{1})(4\omega -\Delta
_{3})(3\omega -\Delta _{3})} \\
&&+\frac{k}{\Delta _{1}\Delta _{3}(\omega -\Delta _{3})}+\frac{c_{0,2}}{%
(2\omega -\Delta _{3})(3\omega -\Delta _{3})} \\
&&\times \left( \frac{c_{0,1}(k+2)}{2\omega -\Delta _{1}}-\frac{c_{1,2}(k+1)%
}{\Delta _{1}}\right) .
\end{eqnarray*}%
We see that for $k>0$ (nonvacuum field states) the CRT are not required for
this effect, but for the photon generation from vacuum either $c_{0,1}$ or
the product $c_{1,2}c_{0,2}$ must be nonzero. In analogy to the generation
of photon pairs in the standard DCE, we call the above effects 3- and
1-photon DCE, respectively.

As seen from Eqs. (\ref{x1}) and (\ref{x2}), to induce the 1- and 3-photon
DCE it is sufficient to modulate just one of the energy levels, yet the
simultaneous modulation of both $E_{1}$ and $E_{2}$ can increase the
transition rate provided the phase difference $(\phi _{1}-\phi _{2})$ is
properly adjusted. However, for constant modulation frequency the photon
generation from vacuum is limited due to the resonance mismatch for
multiphoton dressed states. Indeed, from Eq. (\ref{d}) we have%
\begin{equation*}
\Lambda _{k+J}-\Lambda _{k}=\left( \omega _{ef}+J\alpha \right) J+(2\alpha
J)k\,,
\end{equation*}%
where $J=1,3$. Assuming realistically that $g_{l,k}$ and $\varepsilon _{j}$
are all of the same order of magnitude, we note that $|\alpha |\gtrsim
|\Theta _{k;k+J}^{(\zeta )}|$ for $k\sim 1$. Hence for constant $\eta
_{J}\simeq \Lambda _{J}-\Lambda _{0}$ (adjusted to generate photons from
vacuum) the coupling between the states $|\zeta _{k}\rangle \rightarrow
|\zeta _{k+J}\rangle $ goes off resonance as $k$ increases and we expect a
limited photon production. We note that several methods to enhance the
photon generation were proposed in similar setups, e.g., multi-tone
modulations \cite{jpcs,diego}, time-varying modulation frequency including
effective Landau-Zener transitions \cite{palermo} and optimum control
strategies \cite{optcontrol}.

\section{Discussion and conclusions}

\begin{figure}[tbh]
\centering\includegraphics[width=1.\linewidth]{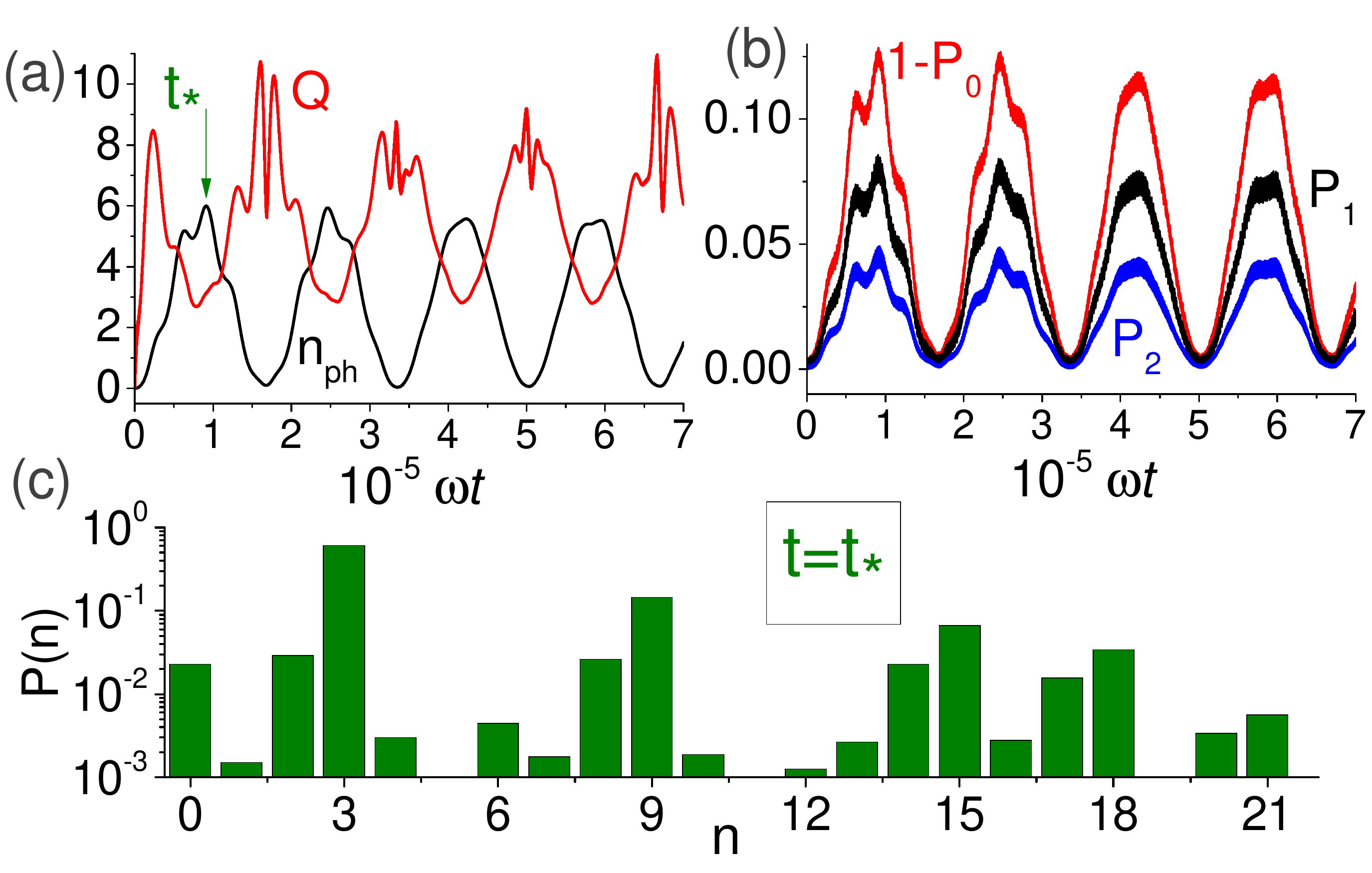}
\caption{(color online) \textbf{System behavior for 3-photon DCE}. a)
Dynamics of the average photon number $n_{ph}$ and the Mandel's $Q$-factor.
b) Dynamics of the atomic populations: the probability that atom leaves the
initial state is $\lesssim 12\%$. c) Photon statistics $P(n)=\mathrm{Tr}(%
\hat{\protect\rho}|n\rangle \langle n|)$ for the time instant $\protect%
\omega t_{\ast }=0.91\times 10^{5}$ [marked by the green arrow in (a)],
where $\hat{\protect\rho}$ is the total density operator. Notice the local
peaks at $n=3k$, asserting the effective 3-photon nature of the process.}
\label{f1}
\end{figure}

\begin{figure}[tbh]
\centering\includegraphics[width=1.\linewidth]{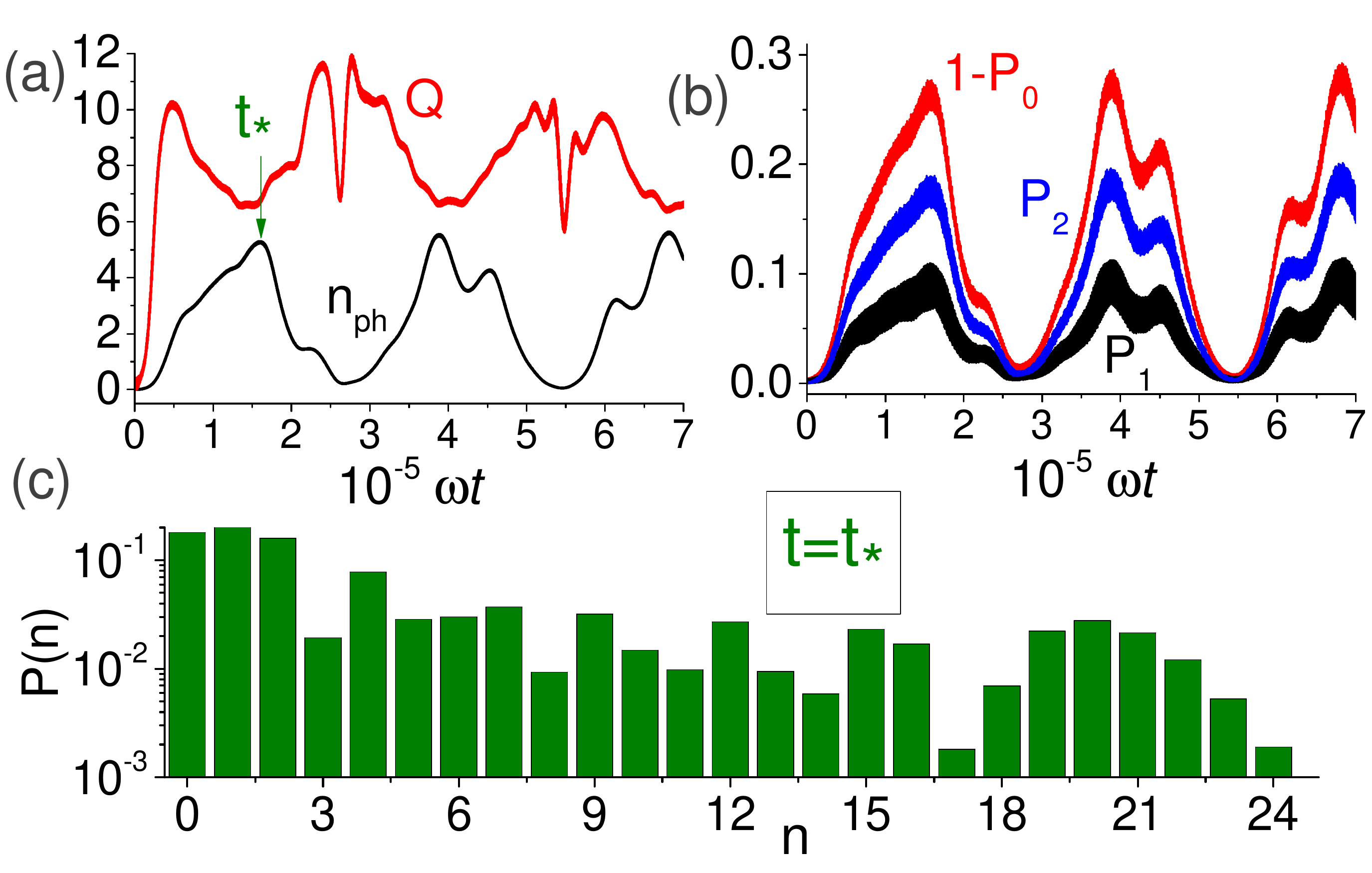}
\caption{(color online) \textbf{System behavior for 1-photon DCE}. Similar
to Fig. \protect\ref{f1}. The probability that the atom leaves the initial
state is now $\lesssim 30\%$. For $\protect\omega t_{\ast }=1.61\times
10^{5} $ (panel c) the photon statistics lacks local peaks, indicating that
the photons are generated via effective 1-photon transitions.}
\label{f2}
\end{figure}

To confirm our analytic predictions we solve numerically the Schr\"{o}dinger
equation for the Hamiltonian (\ref{H2}) considering the initial state $|%
\mathbf{0},0\rangle $ (which is approximately equal to the system ground state in our regime of parameters) and feasible coupling constants $g_{0,1}/\omega
=5\times 10^{-2}$, $g_{1,2}/\omega =6\times 10^{-2}$ and $g_{0,2}/\omega
=3\times 10^{-2}$ (including all the CRT, $c_{l,k}=1$). For the sake of
illustration we consider the sole modulation of $E_{2}$, setting $%
\varepsilon _{1}=0$ and $\varepsilon _{2}=7\times 10^{-2}E_{2}^{(0)}$. In
Fig. \ref{f1} we illustrate the 3-photon DCE for the detunings $\Delta
_{1}/\omega =0.464$, $\Delta _{2}/\omega =0.106$ and modulation frequency $%
\eta /\omega =3.0037$. We show the average photon number $n_{ph}=\langle
\hat{a}^{\dagger }\hat{a}\rangle $, the Mandel's factor $Q=[\langle (\Delta
\hat{n})^{2}\rangle -n_{ph}]/n_{ph}$ (that quantifies the spread of the
photon number distribution, being $Q=1+2n_{ph}$ for the squeezed vacuum
state) and the atomic populations $P_{k}=\langle \hat{\sigma}_{k,k}\rangle $%
. We also show the photon number distribution at the time instant $\omega
t_{\ast }=0.91\times 10^{5}$ (when $n_{ph}$ is maximum), confirming that the
photon generation occurs via effective 3-photon processes. We observe that for $%
t=t_{\ast }$ the photon statistics does not show special behavior around $%
n\approx n_{ph}$. The average photon number and the atomic populations
exhibits a collapse-revival behavior due to increasingly off-resonant
couplings between the probability amplitudes $b_{m}$ in Eq. (\ref{dif}).
Moreover, during the collapses [$n_{ph},(1-P_{0})\approx 0$] the Mandel's
factor is very large, $Q\gg 1,n_{ph}$, which is typical of \emph{hyper-Poissonian}
states that have long tails of distribution with very low (but not
negligible) probabilities \cite{PLAI}.

In Fig. \ref{f2} we perform a similar analysis for the 1-photon DCE, setting
the parameters $\Delta _{1}/\omega =0.362$, $\Delta _{2}/\omega =0.51$ and $%
\eta /\omega =0.9978$. The qualitative behavior of $n_{ph}$, $Q$ and the
atomic populations is similar to the previous case, but the photon number
distribution is completely different, as illustrated in the panel (c) for $%
\omega t_{\ast }=1.61\times 10^{5}$. Now all the photon states are populated
(as expected for an effective 1-photon process), and the $Q$-factor is
always larger that $n_{ph}$ due to the larger spread of the distribution. As
in the previous example, there are no special features in the photon
statistics for $n\approx n_{ph}$, and one has similar probabilities of
detecting any value ranging from 3 to 20 photons.

\begin{figure}[tbh]
\centering\includegraphics[width=1.\linewidth]{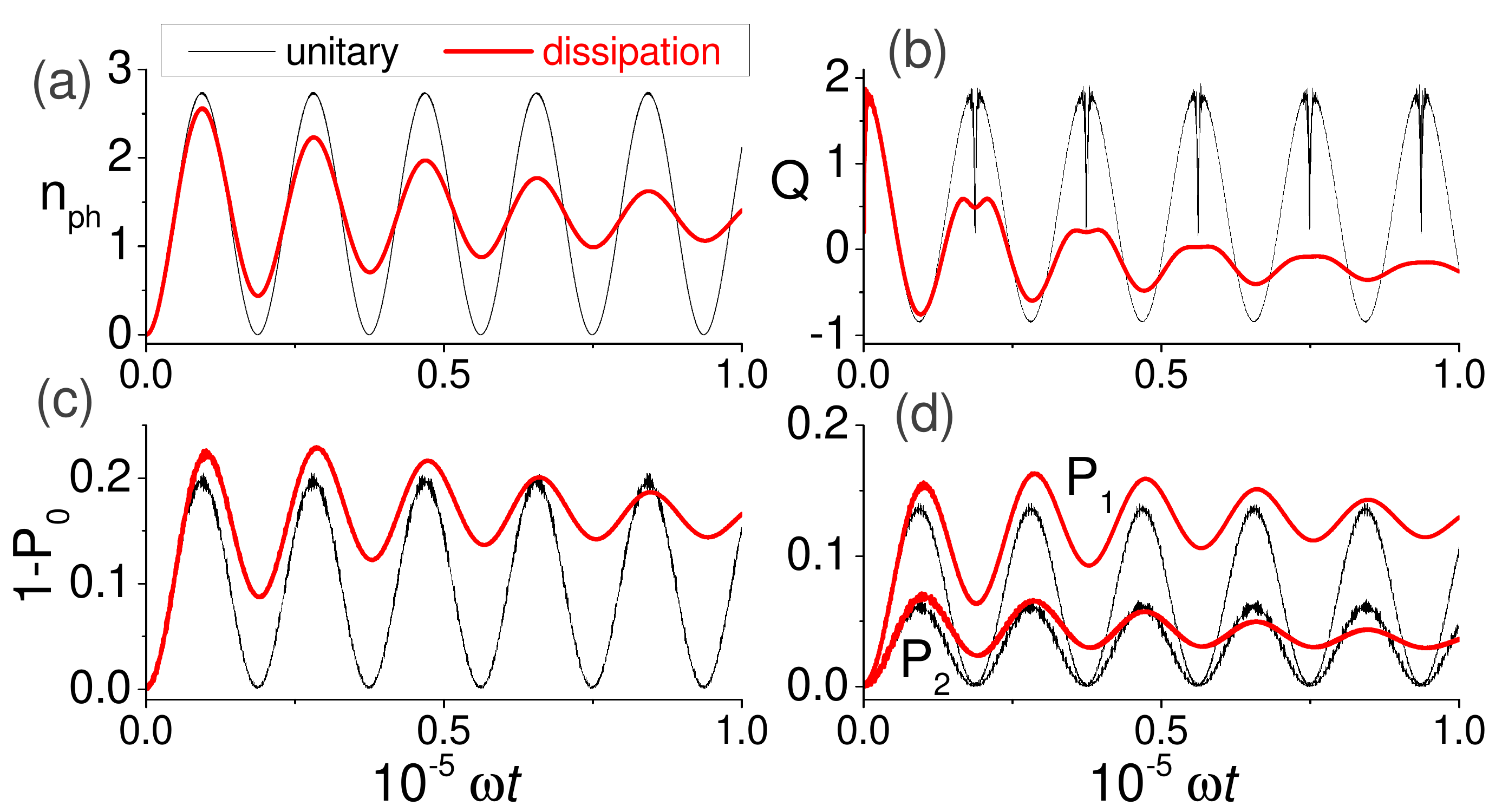}
\caption{(color online) \textbf{Dissipative 3-photon DCE}. Behavior of $%
n_{ph}$, $Q$ and $P_{k}$ under unitary (black thin lines) and dissipative
(red thick lines) evolutions. For initial times (till the first maximum of $%
n_{ph}$) the effects of dissipation are small; for larger times the
dissipation strongly affects the dynamics, but the main qualitative features
persist.}
\label{f3}
\end{figure}

To assess the experimental feasibility of our proposal we solve numerically
the phenomenological Markovian master equation for the density operator $%
\hat{\rho}$ \cite{cycl2,cycl4}%
\begin{equation*}
\dot{\rho}=\frac{1}{i\hbar }[\hat{H},\hat{\rho}]+\kappa \mathcal{L}[\hat{a}%
]+\sum_{k=0}^{1}\sum_{l>k}^{2}\gamma _{k,l}\mathcal{L}[\hat{\sigma}%
_{k,l}]+\sum_{k=1}^{2}\gamma _{k}^{(\phi )}\mathcal{L}[\hat{\sigma}_{k,k}]\,,
\end{equation*}%
where $\mathcal{L}[\hat{O}]\equiv \hat{O}\hat{\rho}\hat{O}^{\dagger }-\hat{O}%
^{\dagger }\hat{O}\hat{\rho}/2-\hat{\rho}\hat{O}^{\dagger }\hat{O}/2$ is the
Lindblad superoperator, $\kappa $ is the cavity relaxation rate and $%
\gamma _{k,l}$ ($\gamma _{k}^{(\phi )}$) are the atomic relaxation (pure
dephasing) rates. Notice that related works demonstrated that for $%
g_{k,l}/\omega < 10^{-1}$ and initial times this approach is a good
approximation to a more rigorous microscopic model of dissipation \cite%
{diego,werlang,palermo}. Typical behavior of 3-photon DCE under unitary and
dissipative evolutions is illustrated in Fig. \ref{f3}, where we set $\Delta
_{1}/\omega =0.24$, $\Delta _{2}/\omega =-0.132$, $\eta /\omega =3.0269$ \cite{solv} and
feasible dissipative parameters $%
\gamma _{k,l}=\gamma _{k}^{(\phi )}=g_{0,1}\times 10^{-3}$ and $\kappa
=g_{0,1}\times 10^{-4}$ (other parameters are as in Fig. \ref{f1}). It is seen
that for initial times the dissipative dynamics resembles the unitary one,
indicating that our predictions could be verified in realistic circuit QED
systems.

In conclusion, we showed that for an artificial cyclic qutrit coupled to a
single-mode cavity one can induce effective 1- and 3-photon transitions
between the system dressed-states in which the atom remains approximately in
the ground state. These effects occur in the dispersive regime of
light-matter interaction for external modulation of some system parameter(s)
with frequencies $\eta \approx \omega $ and $\eta \approx 3\omega $,
respectively. We evaluated the associated transition rates assuming the
modulation of one or both excited energy-levels of the atom, and our method
can be easily extended to the perturbation of all the parameters in the
Hamiltonian. For constant modulation frequency the average photon number and
the atomic populations exhibit a collapse-revival behavior with a limited
photon generation due to effective Kerr nonlinearities. The photon
statistics is strikingly different from the standard (2-photon) DCE case,
for which a squeezed vacuum state would be generated. Although we focused on
transitions that avoid exciting the atom, our approach can be applied to
study other uncommon transitions allowed by $\Delta $-atoms. Hence this
study indicates viable alternatives to engineer effective interactions in
nonstationary circuit QED using cyclic qutrits.

\begin{acknowledgments}
A. V. D. acknowledges a partial support of the Brazilian agency CNPq
(Conselho Nacional de Desenvolvimento Cient\'{\i}fico e Tecnol\'{o}gico).
\end{acknowledgments}

\appendix

\section{Full expressions for the dressed states}\label{apen}

\begin{widetext}

For the purpose of this paper it is sufficient to calculate the eigenstates
of the Hamiltonian $\hat{H}_{0}$ using the second-order perturbation theory.
In the dispersive regime we obtain%
\begin{eqnarray}
|\zeta _{k}\rangle  &=&\mathcal{N}_{k}\left[ |\mathbf{0},k\rangle +\frac{%
g_{0,1}\sqrt{k}}{\Delta _{1}}|\mathbf{1},k-1\rangle -\frac{c_{0,1}g_{0,1}%
\sqrt{k+1}}{2\omega -\Delta _{1}}|\mathbf{1},k+1\rangle -\frac{g_{0,2}%
\sqrt{k}}{\omega -\Delta _{3}}|\mathbf{2},k-1\rangle -\frac{%
c_{0,2}g_{0,2}\sqrt{k+1}}{3\omega -\Delta _{3}}|\mathbf{2},k+1\rangle
\right. \nonumber\\
&&+\left( \frac{c_{0,1}g_{0,1}^{2}}{2\omega -\Delta _{1}}+\frac{%
c_{0,2}g_{0,2}^{2}}{3\omega -\Delta _{3}}\right) \frac{\sqrt{(k+1)(k+2)}%
}{2\omega }|\mathbf{0},k+2\rangle +\left( \frac{c_{0,1}g_{0,1}^{2}}{%
\Delta _{1}}-\frac{c_{0,2}g_{0,2}^{2}}{\omega -\Delta _{3}}\right) \frac{%
\sqrt{k(k-1)}}{2\omega }|\mathbf{0},k-2\rangle  \nonumber\\
&&+\left( \frac{c_{1,2}c_{0,2}(k+1)}{3\omega -\Delta _{3}}+\frac{k}{%
\omega -\Delta _{3}}\right) \frac{g_{1,2}g_{0,2}}{\omega -\Delta _{1}%
}|\mathbf{1},k\rangle +\left( \frac{c_{0,1}(k+1)}{2\omega -\Delta _{1}}-%
\frac{c_{1,2}k}{\Delta _{1}}\right) \frac{g_{0,1}g_{1,2}}{2\omega
_{0}-\Delta _{3}}|\mathbf{2},k\rangle  \nonumber\\
&&+\frac{c_{0,2}g_{1,2}g_{0,2}\sqrt{(k+1)(k+2)}}{(3\omega -\Delta
_{3})(3\omega -\Delta _{1})}|\mathbf{1},k+2\rangle -\frac{%
c_{1,2}g_{1,2}g_{0,2}\sqrt{k(k-1)}}{(\omega -\Delta _{3})(\omega
_{0}+\Delta _{1})}|\mathbf{1},k-2\rangle  \nonumber\\
&&\left. +\frac{c_{0,1}c_{1,2}g_{0,1}g_{1,2}\sqrt{(k+1)(k+2)}}{(2\omega
_{0}-\Delta _{1})(4\omega -\Delta _{3})}|\mathbf{2},k+2\rangle +\frac{%
g_{0,1}g_{1,2}\sqrt{k(k-1)}}{\Delta _{1}\Delta _{3}}|\mathbf{2},k-2\rangle %
\right]  \label{ape},
\end{eqnarray}%
where $\mathcal{N}_{k}=1+O[(g_{0}/\Delta _{1})^{2}]$ is the normalization
constant whose value does not appear in our final (lowest-order) expressions.

For the eigenenergy corresponding to the state $|\zeta _{k}\rangle $ we need
to use the fourth-order perturbation theory to account for the effective
Kerr-nonlinearity. We get%
\begin{equation*}
\Lambda _{k}=\omega k+L_{1}(k)+L_{2}(k)
\end{equation*}%
\begin{equation*}
L_{1}(k)\equiv \left( \delta _{1}-\delta _{2}-c_{0,1}\delta
_{3}-c_{0,2}\delta _{4}\right) k-\left( c_{0,1}\delta _{3}+c_{0,2}\delta
_{4}\right)
\end{equation*}%
\begin{equation*}
L_{2}(k)\equiv \left[ \delta _{1}\beta _{1}(k)-\delta _{2}\beta _{2}(k)%
\right] k-\left[ c_{0,1}\delta _{3}\beta _{3}(k)+c_{0,2}\delta _{4}\beta
_{4}(k)\right] \left( k+1\right) .
\end{equation*}%
We defined the shifts $\delta _{1}=g_{0,1}^{2}/\Delta _{1}$,~$\delta
_{2}=g_{0,2}^{2}/(\omega -\Delta _{3})$,~$\delta
_{3}=g_{0,1}^{2}/(2\omega -\Delta _{1})$,~$\delta
_{4}=g_{0,2}^{2}/(3\omega -\Delta _{3})$,~$\delta
_{5}=g_{1,2}^{2}/(2\omega -\Delta _{3})$,~$\delta
_{6}=g_{1,2}^{2}/(\omega -\Delta _{1})$. Other dimensionless functions
of $k$ are defined as%
\begin{equation*}
\beta _{1}(k)\equiv \left( \delta _{1}-c_{0,2}\delta _{2}\right) \frac{%
c_{0,1}\left( k-1\right) }{2\omega }+\frac{g_{1,2}^{2}\left( k-1\right)
}{\Delta _{1}\Delta _{3}}+c_{1,2}\delta _{5}\left( \frac{c_{0,1}\left(
k+1\right) }{2\omega -\Delta _{1}}-\frac{k}{\Delta _{1}}\right) -\frac{%
L_{1}(k)}{\Delta _{1}}
\end{equation*}%
\begin{equation*}
\beta _{2}(k)\equiv \left( c_{0,1}\delta _{1}-\delta _{2}\right) \frac{%
c_{0,2}\left( k-1\right) }{2\omega }-\frac{c_{1,2}g_{1,2}^{2}\left(
k-1\right) }{\left( \omega -\Delta _{3}\right) \left( \omega +\Delta
_{1}\right) }+\delta _{6}\left( \frac{c_{1,2}c_{0,2}\left( k+1\right) }{%
3\omega -\Delta _{3}}+\frac{k}{\omega -\Delta _{3}}\right) +\frac{%
L_{1}(k)}{\omega -\Delta _{3}}
\end{equation*}%
\begin{equation*}
\beta _{3}(k)\equiv \left( \delta _{3}+c_{0,2}\delta _{4}\right) \frac{k+2}{%
2\omega }+\delta _{5}\left( \frac{\left( k+1\right) }{2\omega
_{0}-\Delta _{1}}-\frac{c_{1,2}k}{\Delta _{1}}\right) +\frac{%
c_{1,2}g_{1,2}^{2}\left( k+2\right) }{\left( 2\omega -\Delta _{1}\right)
\left( 4\omega -\Delta _{3}\right) }+\frac{L_{1}(k)}{2\omega -\Delta
_{1}}
\end{equation*}%
\begin{equation*}
\beta _{4}(k)\equiv \left( c_{0,1}\delta _{3}+\delta _{4}\right) \frac{k+2}{%
2\omega }+c_{1,2}\delta _{6}\left( \frac{c_{1,2}\left( k+1\right) }{%
3\omega -\Delta _{3}}+\frac{k}{\omega -\Delta _{3}}\right) +\frac{%
g_{1,2}^{2}\left( k+2\right) }{\left( 3\omega -\Delta _{3}\right) \left(
3\omega -\Delta _{1}\right) }+\frac{L_{1}(k)}{3\omega -\Delta _{3}}.
\end{equation*}%
\end{widetext}


\begin{thebibliography}{99}
\bibitem{solv} For these parameters $n_{ph}$ is smaller than in Fig. \ref{f1}, so the master equation can be solved numerically by truncating the Fock space at a smaller number of photons.
\end{thebibliography}
\end{document}